*Perspective*

# Physics-Grounded Materials Artificial Intelligence for Reliable Materials Discovery

Yuhang Wang, Qian Wang, Seong-Hoon Jang, and Hao Li*

*Advanced Institute for Materials Research (WPI-AIMR), Tohoku University, Sendai, 980-8577, Japan*

* Corresponding Authors:

Email: li.hao.b8@tohoku.ac.jp (H. L.)

**Abstract**

Artificial intelligence (AI) is transforming materials discovery, yet conventional data-driven approaches often suffer from limited interpretability, poor extrapolation, and inconsistency with physical laws. Since materials behavior is fundamentally governed by thermodynamics, kinetics, electronic structure, transport processes, and operating environments, the next generation of materials intelligence must move beyond correlation-based prediction toward physics-grounded reasoning. In this *Perspective*, we systematically discuss Physics-Grounded Materials AI (PhysMat AI) as a unifying perspective for integrating physical knowledge into materials intelligence through five complementary roles: physics as prior knowledge, descriptors, constraints, verifiers, and infrastructure. Using representative examples from catalysis, solid-state electrolytes in solid-state battery, and hydrogen-storage materials, we illustrate how physical principles guide data representation, model reasoning, validation workflows, and knowledge management. We further present how AI agents can leverage these physics-aware components to perform mechanism-guided discovery within physically feasible search spaces. Finally, we outline a developmental roadmap from physics-aware AI to physics-reasoning AI and ultimately physics-autonomous AI. Looking forward, materials intelligence should evolve from predictive models toward autonomous scientific systems capable of integrating physical reasoning, multiscale simulations, experimental validation, and continuous knowledge updating for reliable materials discovery.

## 1. Introduction

Artificial intelligence (AI) is rapidly transforming the research paradigm of materials science and has demonstrated enormous potential in materials screening, property prediction, structure generation, and experimental optimization.[1-4] In particular, against the backdrop of continued advances in high-throughput computation, automated experimentation, and large-scale materials databases, Materials AI has become an important tool for accelerating materials discovery and design.[5] Over the past decades, Materials AI has evolved from the theory-computation-experiment acceleration framework advocated by the Materials Genome Initiative to recent deep-learning-based materials discovery.[6-7] During this period, mounting evidence has demonstrated substantial improvements in both research efficiency and searchable materials space. For example, Merchant *et al*.[7] developed the GNoME framework by coupling graph neural networks with active learning and, together with high-throughput density functional theory (DFT) calculation validation, achieved large-scale screening of stable crystals, identifying >2.2 million potentially stable structures and substantially expanding both the efficiency and scale of materials discovery. However, despite the remarkable progress achieved by data-driven methods in a number of tasks, reliance on data alone is insufficient to support the next stage of development in materials intelligence.[8] Cheetham *et al*.[9] showed that many AI-predicted inorganic compounds may lack genuine novelty, experimental credibility, and practical utility, as numerous entries appear to be ordered variants of known structures. Their perspective highlights the need to integrate crystallographic and synthetic domain knowledge into AI-driven materials discovery. It should be noted that materials systems are not objects that can be fully described by statistical correlations alone; rather, their behavior is fundamentally governed by the combined effects of thermodynamics, kinetics, electronic structure, interfacial transport, and complex operating environments.[10-13] **Therefore, without support from physical and chemical mechanisms, Materials AI may be able to fit existing data, but it cannot necessarily achieve a true understanding of the governing laws of materials behavior.**

In contrast to fields such as natural language and images, data in materials science are often sparse, heterogeneous, and highly fragmented, and longstanding issues such as small sample size, cross-source bias, and inconsistent label definitions remain pervasive.[14-15] Different studies frequently adopt different experimental conditions, sample preparation protocols, structural characterization methods, and performance evaluation standards, making it difficult to directly align and unify data across sources; even for the same material, the reported properties may be highly sensitive to preprocessing history, defect concentration, testing windows, and environmental variables.[16-17] Moreover, without Findable, Accessible, Interoperable, and Reusable (FAIR) data organization, materials data are often difficult to integrate and reuse owing to limited interoperability across metadata, semantics, and workflows, thereby further complicating cross-source integration and unified representation.[18] More importantly, materials performance is rarely determined by a single label, but instead emerges from the coupling of multiscale and multifactorial effects. In catalytic systems, for example, adsorption behavior, interfacial electric fields, local pH, solvent restructuring, and surface reconstruction/phase transition jointly influence the stability of reaction intermediates, kinetic barriers, and ultimately catalytic selectivity.[10, 19-24] In battery systems, ion migration, interfacial reactions, space charge layers, phase stability, and electrochemical compatibility collectively determine material performance; this is particularly true for solid-state electrolytes (SSEs) in battery applications, in which interfacial stability and cross-interface ion transport often constitute decisive bottlenecks.[12-13] In hydrogen-storage systems, the cooperative relationship among hydrogen adsorption/desorption, bulk diffusion, interfacial mass transport, and dehydrogenation kinetics determines hydrogen storage-release efficiency and cycling stability.[11, 25-26] **Overall, materials problems are never merely problems of data fitting, but scientific problems governed by complex physicochemical mechanisms.**

Against this background, purely black-box AI models face a fundamental challenge: although they may achieve good interpolative predictions within the range of existing data, they often struggle to extrapolate reliably into unknown spaces, especially when the target systems involve new compositional regions, structural prototypes, interfacial states, or operating conditions.[15, 27-28] Models lacking physical constraints are also prone to producing candidate materials that appear highly promising according to scores, yet are in reality not experimentally accessible, unstable, unverifiable, or even inconsistent with basic conservation laws. This not only weakens model interpretability and trustworthiness, but also limits the scientific value of AI in real-world materials discovery. An increasing number of studies have pointed out that machine learning (ML) for materials discovery should not be evaluated solely in terms of predictive accuracy, but must also account for interpretability, uncertainty quantification, extrapolative reliability, mechanistic consistency, and experimental verifiability.[27-30] Therefore, the next stage of Materials AI should not remain at the level of purely data-driven modeling, but should advance toward a new physics-grounded paradigm. **Specifically, future materials intelligence must deeply embed physical laws, chemical principles, and mechanistic understanding into model construction,** **feature representation, inference processes, and validation workflows, enabling AI not only to learn what happens from data, but also to understand why materials behave as they do.**

To systematically address these challenges, a unified framework is required to show how physical knowledge can be integrated into AI across different stages of materials discovery. Inspired by recent advances in physics-informed materials discovery, AI agents, and autonomous scientific research, we organize this *Perspective* around Physics-Grounded Materials AI (PhysMat AI). The core of PhysMat AI is to integrate physical constraints, domain knowledge, mechanistic descriptors, database systems, and experimental feedback into a unified framework. This framework transforms AI from a purely data-driven learner into an intelligent discovery system with physical awareness, mechanistic reasoning, and closed-loop validation capabilities. From this view, this *Perspective* systematically discusses the roles of physics in catalysis, SSEs in solid-state battery, and hydrogen-storage materials. **We demonstrate how physics functions as prior knowledge, descriptors, constraints, verifiers, and infrastructure within PhysMat AI. Based on these foundations, we further propose a developmental roadmap that progresses from physics-aware AI to physics-reasoning AI and ultimately to physics-autonomous AI.** We emphasize that the future of materials AI is not black-box automation detached from scientific laws, but an autonomous intelligent discovery system built upon physical foundations.

## 2. Understanding PhysMat AI

PhysMat AI is a physics-guided materials AI framework that enables reliable, interpretable, and verifiable materials discovery. **Figure 1** presents the overall workflow of PhysMat AI, illustrating how physical knowledge, databases, AI models, validation, and feedback are integrated into a closed-loop materials discovery process. First, physical laws and theoretical principles provide reliable prior knowledge for AI models, ensuring that the learning process remains consistent with fundamental scientific rules. Second, high-quality materials databases are established by integrating multimodal data sources, including experimental measurements, computational results, and scientific literature, thereby providing a robust foundation for model training. Building upon this foundation, ML models, large language models (LLMs), and multi-agent systems are employed to perform property prediction, mechanism understanding, and candidate screening. Subsequently, AI-generated predictions are verified through theoretical calculations, experimental characterization, and performance evaluation. Finally, the validated results are fed back into the databases and AI models, enabling continuous knowledge updating and iterative model refinement. This closed-loop framework facilitates the transition from purely data-driven prediction to physics-guided discovery,[31] providing a unified framework for advancing catalysis, energy storage, hydrogen storage, and autonomous scientific research systems.

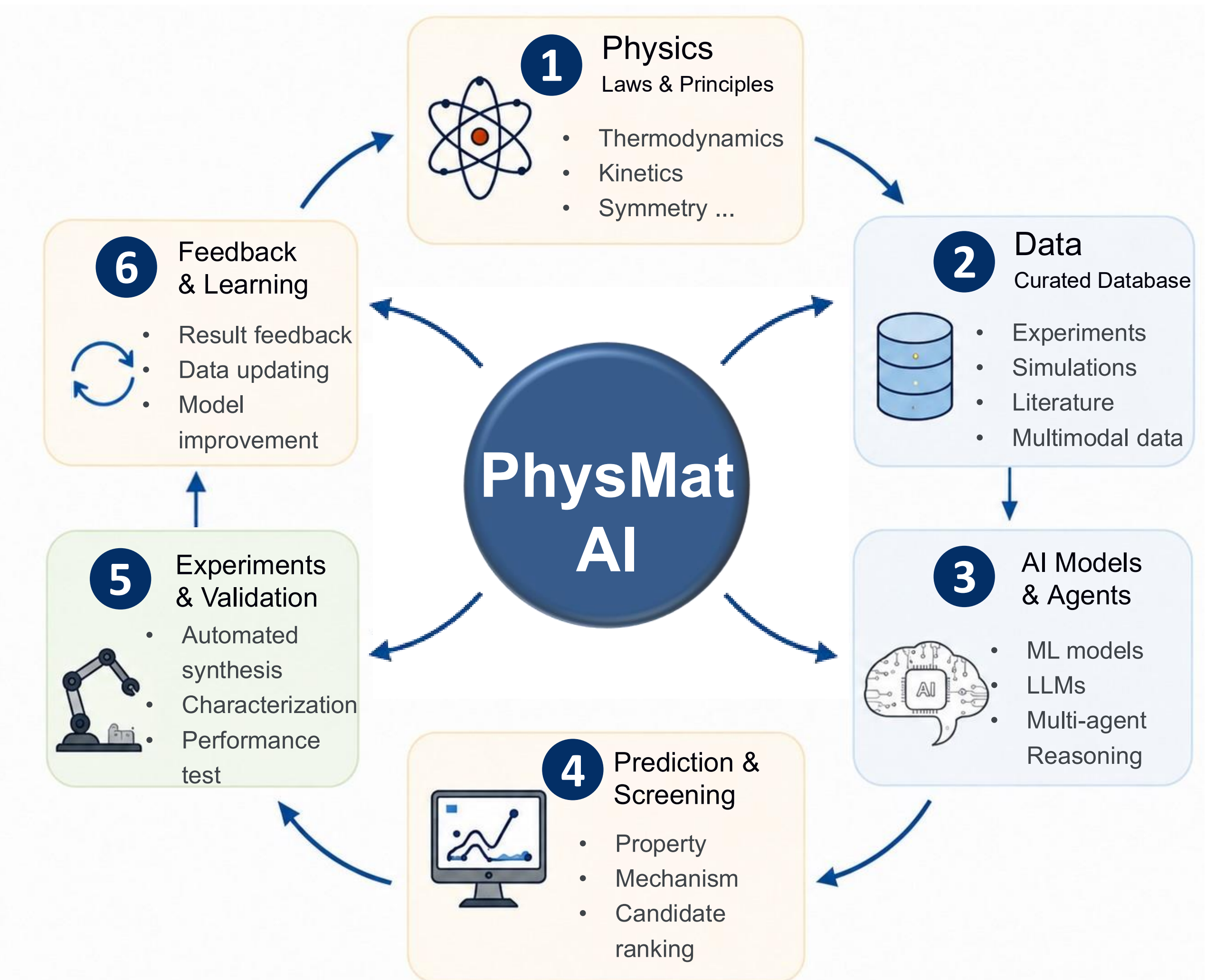

**Figure 1.** Overall closed-loop workflow of Physics-Grounded Materials AI (PhysMat AI). It describes a closed-loop materials-discovery workflow that integrates physical principles, curated databases, AI models and agents, prediction and screening, experimental validation, and continuous feedback. Physics provides the scientific foundation for data representation and model reasoning, while validated results are continuously fed back to improve both the knowledge base and AI models, enabling reliable, interpretable, and continuously evolving materials discovery.

While **Figure 1** focuses on the overall discovery workflow, **Figure 2** further reveals the internal five-layer physics architecture that enables each stage of this workflow. Unlike conventional materials AI systems that primarily rely on data-driven learning, PhysMat AI achieves a deep integration of physical knowledge and AI through a five-layer framework. First, **Physics as Prior** embeds physical laws and theoretical mechanisms into AI models as prior knowledge. Next, **Physics as Descriptor** transforms complex physical processes into learnable descriptors and features. Building upon these representations, **Physics as Constraint** incorporates physical principles to guide model reasoning and decision-making. Subsequently, **Physics as Verifier** combines theoretical calculations and experimental validation to assess the reliability of AI-generated predictions. Finally, **Physics as Infrastructure** establishes the data foundation that supports model training, knowledge accumulation, and continuous optimization. Together, these five layers collectively span the full lifecycle of materials intelligence, from knowledge encoding and representation to reasoning, validation, and knowledge accumulation.

Layer 1
Physics as Prior
Physical Knowledge Layer
Physics-informed prior knowledge
Laws · Principles · Theories · Mechanisms
Embed physical understanding as AI prior
Layer 2
Physics as Descriptor
Representation Layer
Physics-derived descriptors
Mechanism-informed features · Descriptors · Embeddings
Translate physics into learnable representations
Layer 3
Physics as Constraint
Reasoning Layer
Physics-constrained reasoning
AI models & agents · Mechanism reasoning · Decision making
Guide learning and prediction with physical constraints
Layer 4
Physics as Verifier
Verification Layer
Physics-verified closed loop
Theory calculations . Experiments Validation Feedback
Verify predictions and refine knowledge
Layer 5
Physics as Infrastructure
Data Layer
Physics-aware data infrastructure
Curated data · Simulations . Literature · Metadata
Provide reliable, standardized, and scalable data foundation

**Figure 2.** Five-layer physics architecture underlying PhysMat AI. The five-layer architecture illustrates how physical knowledge is systematically integrated into materials intelligence through prior knowledge, descriptors, constraints, verification, and infrastructure. These complementary layers

span the complete AI workflow, from knowledge representation and physics-guided reasoning to validation and data infrastructure, providing a unified perspective for reliable, interpretable, and continuously evolving materials discovery.

## 3. The Roles of Physics in PhysMat AI across Catalysis, Solid-State Battery, and Hydrogen Storage

Energy conversion and storage have long been among the central themes of materials science, encompassing diverse yet fundamentally connected fields such as catalysis, solid-state battery (particularly SSEs), and hydrogen storage.[2] Despite differences in target functions and operating conditions, these systems are governed by common physical principles, including thermodynamics, kinetics, electronic structure, interfacial phenomena, and mass transport. Therefore, they provide ideal testbeds for examining how physical knowledge can be systematically integrated into AI. Through representative examples from catalysis, SSEs, and hydrogen storage, we show that physics is not merely auxiliary information for AI models or agents, but a fundamental component that shapes data representation, model reasoning, scientific interpretation, and autonomous materials discovery.

### 3.1 Physics as Prior: Embedding Physical Principles into Materials Intelligence

#### 3.1.1 General Description

The first layer incorporates established physical and chemical knowledge into AI models, thereby improving model reliability, interpretability, and extrapolation capability.[32] **Figure 3a** illustrates how physical knowledge can be incorporated into AI models as prior information, rather than being inferred solely from data-driven learning. Within this framework, domain knowledge such as conservation laws, stability principles, scaling relationships, reaction kinetics, and interfacial effects is encoded as knowledge modules and integrated with materials data as inputs to the model. Compared with conventional black-box learning approaches, physical priors can effectively reduce the search space while improving model generalization and physical consistency. By incorporating domain-specific knowledge, AI systems can not only capture statistical correlations among material properties but also establish representations and reasoning processes that are aligned with underlying physical mechanisms. As a result, PhysMat AI enables more reliable, interpretable, and transferable materials discovery with enhanced extrapolation capability beyond the available data.

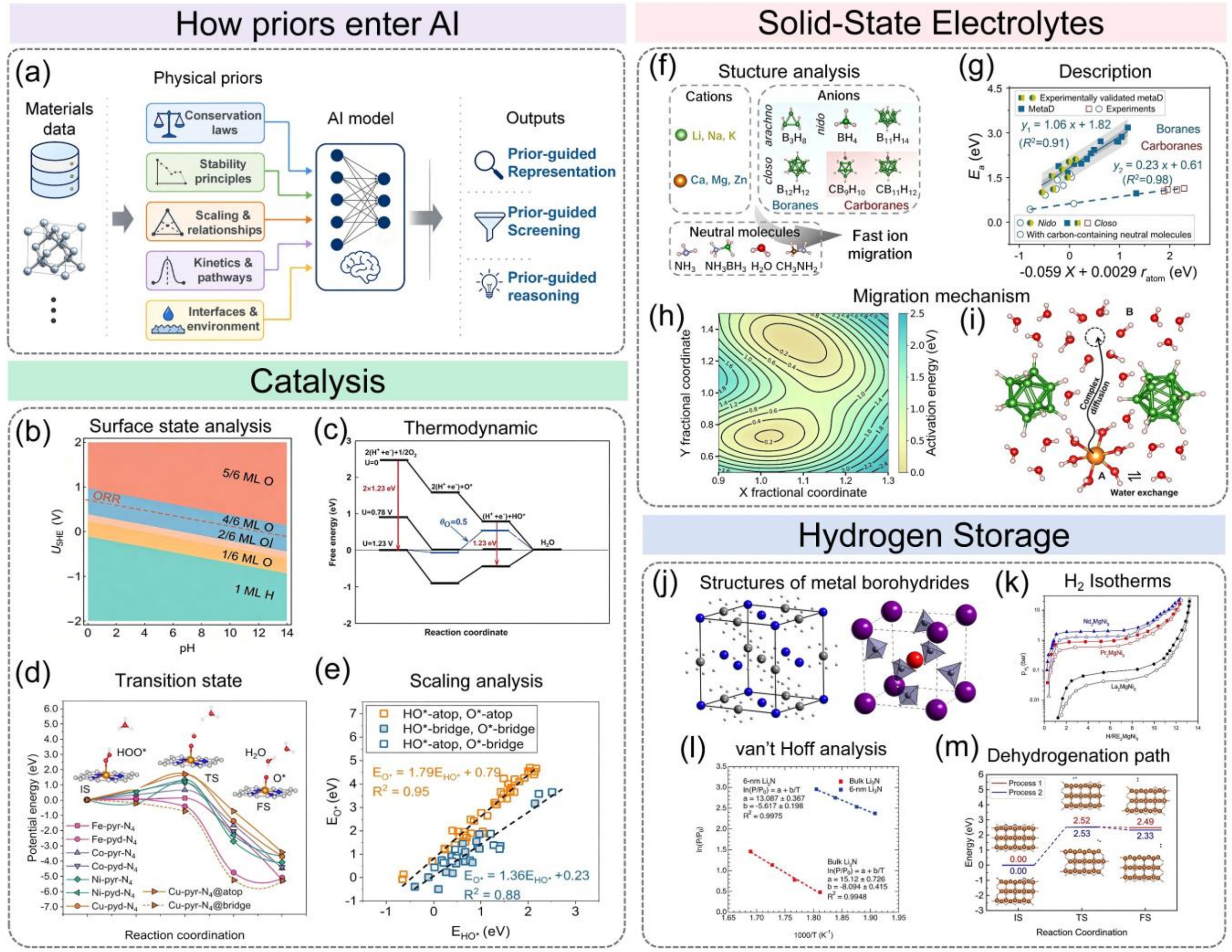

**Figure 3**. (a) Conceptual illustration of Physics as Prior in PhysMat AI. **Catalysis:** (b) 2D surface Pourbaix diagram for ZrN(100) as an example. Reproduced with permission from ref.[33] licensed under a Creative Commons License CC BY-NC 3.0. (c) Free-energy diagram for oxygen reduction over Pt(111) as an example. Reproduced with permission from ref.[34] copyright 2024, American Chemical Society. (d) Transition-state energy barriers for O-O bond cleavage on M-N-C catalysts. (e) Scaling relations between binding energies $E_{O*}$ and $E_{HO*}$. Reproduced with permission from ref.[35] licensed under a Creative Commons License CC BY-NC 4.0. **Solid-state electrolytes:** (f) Typical cations, anions, and neutral molecules in hydride SSEs. (g) Multiple linear regressions for divalent SSEs with neutral molecules. (h) Potential energy surfaces of Mg-ion migration. Reproduced with permission from ref.[36] licensed under a Creative Commons License CC BY-NC 4.0. (i) Illustration of the migration path of the $[Mg(H_2O)_x]^{2+}$ hydro complex. Reproduced with permission from ref.[37] copyright 2023, American Chemical Society. **Hydrogen Storage:** (j) Crystal structures of selected

metal borohydrides, including $NaBH_4$ and $CsSr(BH_4)_3$. Reproduced with permission from ref.[38] licensed under a Creative Commons License CC BY-NC 4.0 and ref.[39] copyright 2016, Royal Society of Chemistry. (k) Evolution of isotherms and equilibrium pressures of H absorption and desorption at room temperature. Reproduced with permission from ref.[40] copyright 2015, Elsevier. (l) van't Hoff analysis for determining thermodynamics of hydrogen desorption. Reproduced with permission from ref.[41] copyright 2024, Wiley-VCH. (m) Surface dehydrogenation calculated using the climbing image nudged elastic band method. Reproduced with permission from ref.[42] copyright 2022, Royal Society of Chemistry.

#### 3.1.2 Examples in Catalysis, SSEs, and Hydrogen Storage

*In catalysis*, AI does not autonomously discover all governing principles from raw data; rather, it is built upon a well-established theoretical framework developed over decades of catalytic research. **Figures 3b-e** present several representative knowledge modules that constitute the physical priors of catalytic AI.

First, surface Pourbaix diagrams describe the thermodynamically stable surface states of electrocatalysts under different potentials and pH conditions (**Figure 3b**), providing a fundamental basis for identifying realistic active sites. Notably, electrochemistry-driven pre-adsorbed molecules generated from the liquid phase may either modify the electronic properties of the catalyst surface or even block active sites.[33, 43] Second, free-energy diagrams provide a theoretical framework for identifying reaction pathways and rate-limiting steps, thereby elucidating the origins of catalytic activity. The computational hydrogen electrode (CHE) approach proposed by Nørskov *et al*.[34] established a thermodynamic framework to relate DFT-calculated adsorption energies to electrode potentials (**Figure 3c**). Furthermore, transition-state analysis characterizes reaction kinetics through activation barriers and establishes direct links between catalyst structure and reaction rates (**Figure 3d**). Unlike thermodynamic analyses that rely solely on the energies of initial and final states, transition-state calculations identify the kinetically rate-determining steps and reveal the actual kinetic bottlenecks. Finally, scaling relationships describe the intrinsic linear correlations among adsorption energies of reaction intermediates.[35, 44-45] The physical origin of these relationships lies in the fact that these intermediates interact with the catalyst surface through the same active site and similar bonding mechanisms.[46] By linking a small number of readily accessible adsorption descriptors to the thermodynamics and, in many cases, the kinetics of catalytic reactions, scaling relationships provide a unified framework for predicting reaction energetics and catalyst performance (**Figure 3e**).

Collectively, these theoretical concepts constitute the physical priors of catalytic AI. By embedding established catalytic principles into the learning process, AI models can perform prediction and discovery while adhering to fundamental catalytic laws, rather than relying solely on statistical correlations in data.

*For SSEs in solid-state battery*, the primary challenge is that ionic conductivity is not a direct compositional property but an emergent consequence of ion migration within a complex structural and electrochemical environment.[47] Fast ion transport depends on defect chemistry, lattice dynamics, local disorder, migration-network connectivity, and interfacial stability, while practical implementation additionally requires electronic insulation, electrochemical robustness, mechanical integrity, and compatibility with electrode materials.[48-50] Consequently, SSE discovery is fundamentally a physics-constrained optimization problem rather than a simple search for high-conductivity compositions. As illustrated in **Figure 3f**, SSEs contain a diverse range of mobile cations, complex anion frameworks, and neutral molecular species, all of which collectively determine ion transport behavior. The resulting migration characteristics emerge from the interplay between local coordination environments, structural disorder, and framework dynamics rather than from composition alone.

This distinction is important for AI-guided discovery. Traditional ML workflows often treat conductivity as a target variable and search for statistical relationships between composition and performance. However, identical conductivity values may originate from very different transport mechanisms.[51] In garnet electrolytes, Li transport is strongly influenced by Li-site occupancy, vacancy concentration, and migration-network connectivity, whereas in argyrodite electrolytes activation barriers are closely associated with lattice expansion, bottleneck size, and anion disorder.[52-53] Similar transport–structure relationships have been reported in NASICON and sulfide electrolytes, where framework flexibility and local structural heterogeneity govern ion mobility.[54-55] Such transport–structure relationships can often be quantified through physically meaningful relationships. For example, as shown in **Figure 3g**, activation energies in hydride electrolytes exhibit strong correlations with atomistic features derived from local structural environments, highlighting how physically interpretable features can capture the underlying mechanisms governing ionic transport.

These observations suggest that AI systems should begin from mechanistic questions rather than performance labels. Instead of asking which composition yields the highest conductivity, a physics-guided framework seeks to understand how ions migrate, which structural features facilitate transport, and which degradation pathways limit performance. As illustrated in **Figure 3h**, the migration-energy landscape and atomistic diffusion pathway reveal how local coordination environments, migration bottlenecks, and energy barriers collectively govern ionic transport. Such mechanistic understanding provides the physical prior that guides data representation, model construction, and interpretation, allowing AI to learn transport mechanisms rather than merely fitting conductivity values. Overall, physical transport mechanisms can serve as the prior knowledge that enables AI to discover SSEs within physically meaningful chemical and transport spaces.

*For hydrogen-storage materials*, practical performance cannot be evaluated by storage capacity alone. A viable material must satisfy multiple coupled requirements, including hydrogen-storage density, equilibrium pressure, operating temperature, kinetics, and cycling reversibility. These properties are governed by thermodynamics, phase transformations, diffusion, and interfacial processes. Consequently, AI should reason over physically meaningful operating windows rather than simply predicting capacity from composition.[56-62] As illustrated in **Figure 3j**, hydrogen-storage materials encompass diverse crystal structures and hydrogen-hosting chemistries. These structural characteristics provide the fundamental physical basis for hydrogen accommodation and determine the subsequent thermodynamic and kinetic behavior.

Thermodynamic priors are essential because pressure–composition–temperature (PCT) behavior determines whether a hydride can operate within a desired pressure–temperature range.[11] **Figure 3k** presents representative hydrogen absorption/desorption isotherms, while **Figure 3l** shows the corresponding van't Hoff analysis used to extract thermodynamic parameters. Together, they establish the relationship between equilibrium pressure, temperature, and hydrogenation thermodynamics, defining the practical operating window of hydrogen-storage materials. Plateau pressures, hysteresis, PCT/van't Hoff-derived hydrogenation enthalpy and entropy, and phase-equilibrium boundaries associated with hydrogen-induced phase transitions define the thermodynamic boundary between hydrogen uptake and release. In this sense, the van't Hoff relation is not merely a fitting tool applied after measuring PCT curves; **it defines a physical condition that the model must respect**. A high-capacity material with an equilibrium pressure far outside the intended operating window is not useful, even if its predicted storage density is large. Conversely, a material with moderate capacity may become attractive if it combines an appropriate plateau pressure with fast kinetics and stable cycling.[59-61, 63-65]

Kinetic priors prevent AI from confusing thermodynamic possibility with accessible performance. $MgH_2$ is a representative example. As shown in **Figure 3m**, atomistic calculations reveal the dehydrogenation pathway and the associated reaction barriers, providing mechanistic insight into the kinetic processes governing hydrogen release. Although $MgH_2$ offers high gravimetric hydrogen capacity, its dehydrogenation behavior is controlled by surface desorption, nucleation, phase transformation, product-layer growth, and bulk diffusion.[62, 66] Atomistic and data-driven studies show that early-stage hydrogen release can involve a "burst-like" (or "dam-break") effect, where the release of surface hydrogen is usually the kinetically limiting step.[42, 67] Such multi-regime behavior means that desorption curves should be read as physical trajectories, not merely as smooth labels for onset temperature or activation energy. Particle size, catalyst distribution, defect density, surface conditions, product-layer morphology, heat transport, and cycling history all shape the observed release response.[42, 60-64, 66-67]

Rather than optimizing a single property, AI should identify materials that satisfy coupled requirements of capacity, thermodynamics, kinetics, and reversibility within practical operating conditions. Such mechanistic understanding provides the physical foundation that guides data representation, model construction, and materials discovery from the outset.

## 3.2 Physics as Descriptor: Translating Physical Mechanisms into Learnable Representations

### 3.2.1 General Description

The central idea of this layer is to transform complex physical mechanisms into feature representations that can be understood and learned by AI models. As illustrated in **Figure 4a**, raw materials data often lack explicit physical meaning and therefore require a descriptor extraction process. This process maps key physical quantities such as structural information, electronic properties, thermodynamic characteristics, kinetic parameters, and environmental factors into a learnable feature space. Compared with directly training models on raw data, physics-informed descriptors can significantly improve both model interpretability and generalization by establishing explicit connections between the learning process and the underlying physical mechanisms.

Through this transformation, physical knowledge is converted into structured representations that serve as a critical bridge between materials science and AI. These representations provide the foundation for subsequent tasks such as property prediction, mechanism understanding, and materials design. Importantly, descriptors can be derived from microscopic information, including atomic configurations and electronic structures, as well as from macroscopic physical principles such as thermodynamics, kinetics, and environmental effects. Together, these descriptor representations provide a unified means of translating domain knowledge into machine-learnable features while retaining clear physical meaning across different materials systems.

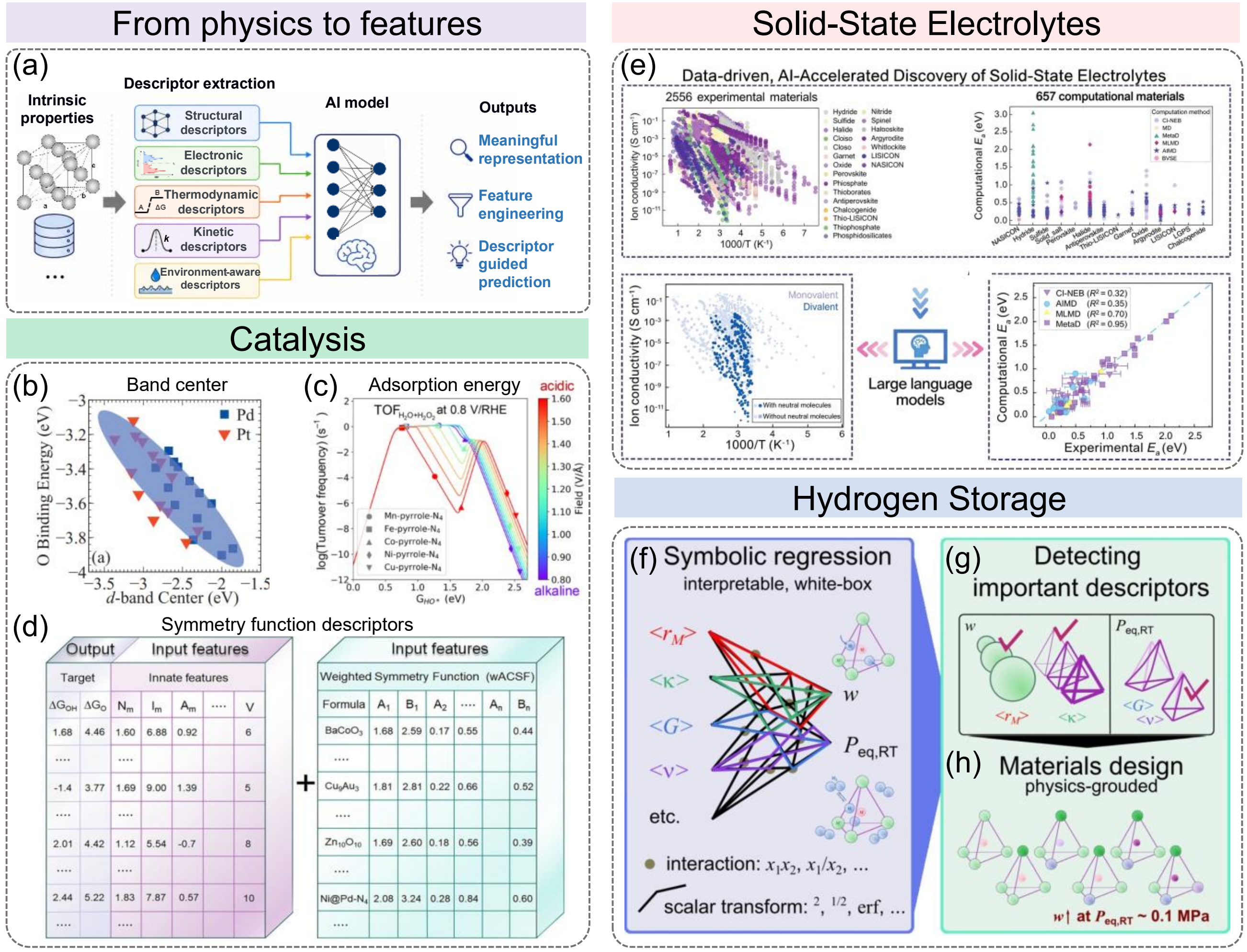

**Figure 4.** (a) Conceptual illustration of physics as descriptor in PhysMat AI. **Catalysis:** (b) Adsorption energies of O and OH on Pd and Pt skin alloys as a function of *d* band center. Reproduced with permission from ref.[68] copyright 2010, American Institute of Physics. (c) Activity volcano for ORR of M-pyrrole-N catalysts. Reproduced with permission from ref.[69] licensed under a Creative Commons License CC BY-NC 4.0. (d) Feature selection *via* Pearson correlation and recursive elimination based on weighted atomic-centered symmetry function (wACSF) models. Reproduced with permission from ref.[70] licensed under a Creative Commons License CC BY-NC 4.0. **Solid-state electrolytes:** (e) Upper: Ionic conductivity and activation-energy distributions of SSEs. Lower left: Temperature-dependent ionic conductivity of hydride electrolytes. Lower right: Benchmark comparison between experimental and computational activation energies. Reproduced with permission

from ref.[36] licensed under a Creative Commons License CC BY-NC 4.0. **Hydrogen storage:** (f) Interpretable symbolic regression linking descriptors to hydrogen-storage performance metrics. (g) Key descriptors governing hydrogen-storage performance in interstitial hydrides. (h) Descriptor-based design principles for balancing storage capacity and equilibrium pressure. Reproduced with permission from ref.[71] licensed under a Creative Commons License CC BY-NC 3.0.

### 3.2.2 Examples in Catalysis, SSEs and Hydrogen Storage

*In catalysis*, the primary objective is to transform complex catalytic mechanisms into quantifiable and learnable physical descriptors, thereby establishing explicit relationships between material structures and catalytic performance.

Electronic-structure-related descriptors have been extensively applied in catalytic research.[72] **Figure 4b** illustrates the classical *d* band theory, in which adsorption energies exhibit a strong correlation with the *d* band center (*i.e.*, the average energy of *d* electrons), reflecting the fundamental role of the electronic structure of transition-metal surfaces in regulating adsorption strength.[73] According to the Hammer-Nørskov model,[74] the adsorption energy can be approximately described as the result of competition between the coupling of adsorbate states with metal *d* states and Pauli repulsion. Consequently, the *d* band center has become one of the most widely used electronic descriptors for correlating catalytic activity.[73, 75] Building upon electronic descriptors, **Figure 4c** illustrates the general concept of descriptor-based volcano relationships, in which the adsorption free energy of key reaction intermediates serves as a physically meaningful descriptor for correlating catalytic activity.[69]

Furthermore, recent studies have demonstrated that factors such as magnetic field effects,[76-78] as well as dipole moments and polarizability,[19] can shift the position of the volcano peak and alter catalytic activity trends. These findings suggest that catalyst descriptors can be extended beyond static adsorption energies to include physically meaningful quantities directly associated with operating conditions, thereby enabling a more realistic description of catalytic behavior under practical reaction environments.

Beyond conventional descriptors, **Figure 4d** illustrates a high-dimensional descriptor framework based on weighted atom-centered symmetry functions (wACSF).[70] By jointly encoding local coordination environments, geometric structures, and chemical attributes, wACSF provides a physically meaningful representation of catalytic active sites. When combined with elemental properties such as electronegativity and valence electron count, it establishes a transferable feature space that enables AI models to capture complex structure-activity relationships.

Overall, descriptor development has evolved from simple electronic descriptors, such as the *d*-band center, to adsorption-energy descriptors and ultimately to high-dimensional representations such as wACSF. This progression enables AI models to capture increasingly complex catalytic environments while preserving physical interpretability, providing a robust foundation for catalyst design and accelerated materials discovery.

*For SSEs in solid-state battery*, effective descriptors should not be limited to elemental composition or reported room-temperature conductivity, because these quantities often do not directly encode the microscopic origin of ion transport. Instead, descriptors should capture transport-relevant physics, including migration barriers, bottleneck size, local coordination environments, site occupancy, defect concentration, lattice softness, structural disorder, diffusion-network connectivity, electrochemical stability, and mechanical properties.

This descriptor logic is already evident in atomistic modeling studies. DFT can provide migration barriers, defect formation energies, phase stability, and electrochemical stability windows, while *ab initio* molecular dynamics (AIMD) can reveal finite-temperature diffusion, correlated hopping, and local structural rearrangement. However, ionic transport is a rare-event process governed by thermally activated hopping and collective ion motion, which are often difficult to sample using short AIMD trajectories.[52] This limitation motivates the use of machine learning interatomic potential (MLIP), which extend near-first-principles-level simulations to longer time scales and larger length scales while retaining an atomistic description of the potential energy surface.[79-81]

MLIP-based simulations further expand the descriptor space from static structural features to dynamic transport features. Correlated hopping topology, migration network connectivity, lattice softness, disorder-mediated transport, and finite-size-aware diffusion statistics can serve as physically meaningful descriptors for collective ion migration.[82-83] For example, in argyrodite electrolytes, lattice expansion and anion disorder modify bottleneck geometries and directly affect Li-ion activation barriers.[53] In garnet systems, aliovalent doping alters Li-site occupancy and vacancy distributions, thereby reshaping the migration network.[84] Such descriptors provide mechanistic information that can be compared across materials classes and integrated into autonomous discovery workflows.

The descriptor-centric nature of SSE discovery is illustrated in **Figure 4e**. Large-scale experimental and computational datasets reveal substantial variations in ionic conductivity and activation energy across different electrolyte families. Rather than relying solely on composition, physically meaningful descriptors such as activation barriers, transport-related structural features, and local coordination environments provide a more direct connection to ion-transport mechanisms. The strong agreement between experimentally measured and computationally predicted activation energies further demonstrates the potential of physics-informed descriptors for data-driven SSE discovery.

Across different materials systems, descriptors consistently translate complex physical mechanisms into transferable representations that improve both model interpretability and generalization. By integrating atomistic simulations, transport physics, and data-driven learning, these descriptors provide interpretable connections between local structures and transport properties, supporting reliable prediction, mechanistic understanding, and the discovery of next-generation SSEs.

*In hydrogen-storage*, descriptors bridge physical mechanisms and AI models by translating hydrogen uptake, thermodynamic stability, and transport behavior into quantitative representations that can be learned, interpreted, and generalized across different materials systems. Rather than relying solely on elemental composition, AI models should encode the mechanisms governing hydrogen-storage performance, including host-lattice accommodation, hydride stability, elastic response, surface chemistry, and hydrogen diffusion.[71, 85-86] In this sense, descriptors represent compressed physical hypotheses that transform mechanistic understanding into learnable representations.

Recent studies demonstrate that compact and physically interpretable descriptor sets can achieve predictive performance comparable to black-box models while preserving chemical transparency.[86] Atomic mass, density, and lattice descriptors primarily capture gravimetric and volumetric storage characteristics, whereas electronegativity- and elasticity-related descriptors reflect hydride stability, equilibrium pressure, and structural accommodation.[71, 85-86] Importantly, the unified descriptor framework for interstitial hydrides separates storage capacity from equilibrium pressure by associating them with distinct physical mechanisms, preventing AI models from collapsing hydrogen storage into a single hidden score.[72]

As illustrated in **Figure 4f-h**, descriptor engineering can be further integrated with interpretable AI workflows. White-box symbolic regression identifies explicit descriptor interactions without sacrificing physical interpretability, while feature-selection methods reveal the descriptors that dominate storage capacity and room-temperature equilibrium pressure.[71, 87] These physically informed descriptors are subsequently translated into mechanism-based design principles, enabling AI to distinguish competing hydrogen-storage mechanisms and guide materials optimization. Kinetic descriptors provide an additional layer of representation by incorporating activation barriers, diffusion coefficients, migration-network connectivity, and interface characteristics, whereas MLIP and finite-temperature simulations further extract dynamic descriptors associated with hydrogen mobility and defect-sensitive transport.[88-90]

In summary, mechanism-aware descriptor representations allow AI models to move beyond composition-based prediction by capturing the thermodynamic, kinetic, and structural origins of hydrogen storage performance, thereby supporting interpretable materials discovery and rational design.

### 3.3 Physics as Constraint: Physics-Guided Reasoning and *Operando* Constraints

The third core layer of PhysMat AI is that physics is not only embedded as prior knowledge in the input but also directly constrains the prediction space and reasoning behavior of AI models, enabling physics-guided reasoning, namely physics inside the model. Physical laws restrict AI reasoning to thermodynamically and kinetically feasible regions of the search space.[91]

This layer emphasizes that physical laws are not only used to construct descriptors and features but also actively participate in the learning, reasoning, and decision-making processes of AI models. As illustrated in **Figure 5a**, conventional data-driven models typically search for optimal solutions within an extremely large design space. **However, a high prediction score does not necessarily imply physical feasibility.** In materials systems, many predicted candidates may suffer from thermodynamic instability, kinetically inaccessible reaction pathways, phase instability, or a lack of experimental synthesizability. Consequently, models that rely solely on statistical correlations are prone to generating predictions that appear highly promising but are physically unrealistic. The fundamental reason is that correlation does not necessarily imply physical feasibility.

To overcome this limitation, AI models incorporate multiple levels of physical guidance that progressively restrict the search space to physically accessible regions (**Figure 5b**). These constraints may originate from thermodynamics, kinetics, phase stability, transport behavior, operating conditions, and uncertainty quantification. Rather than searching the entire design space, AI models are guided toward candidate materials that satisfy fundamental physical requirements. In practice, these constraints can be implemented as sequential feasibility filters, physics-based objective functions, or tool-assisted reasoning steps. Rather than optimizing a prediction score alone, AI systems progressively eliminate candidates that violate thermodynamic stability, kinetic accessibility, phase stability, or operating constraints before further evaluation.

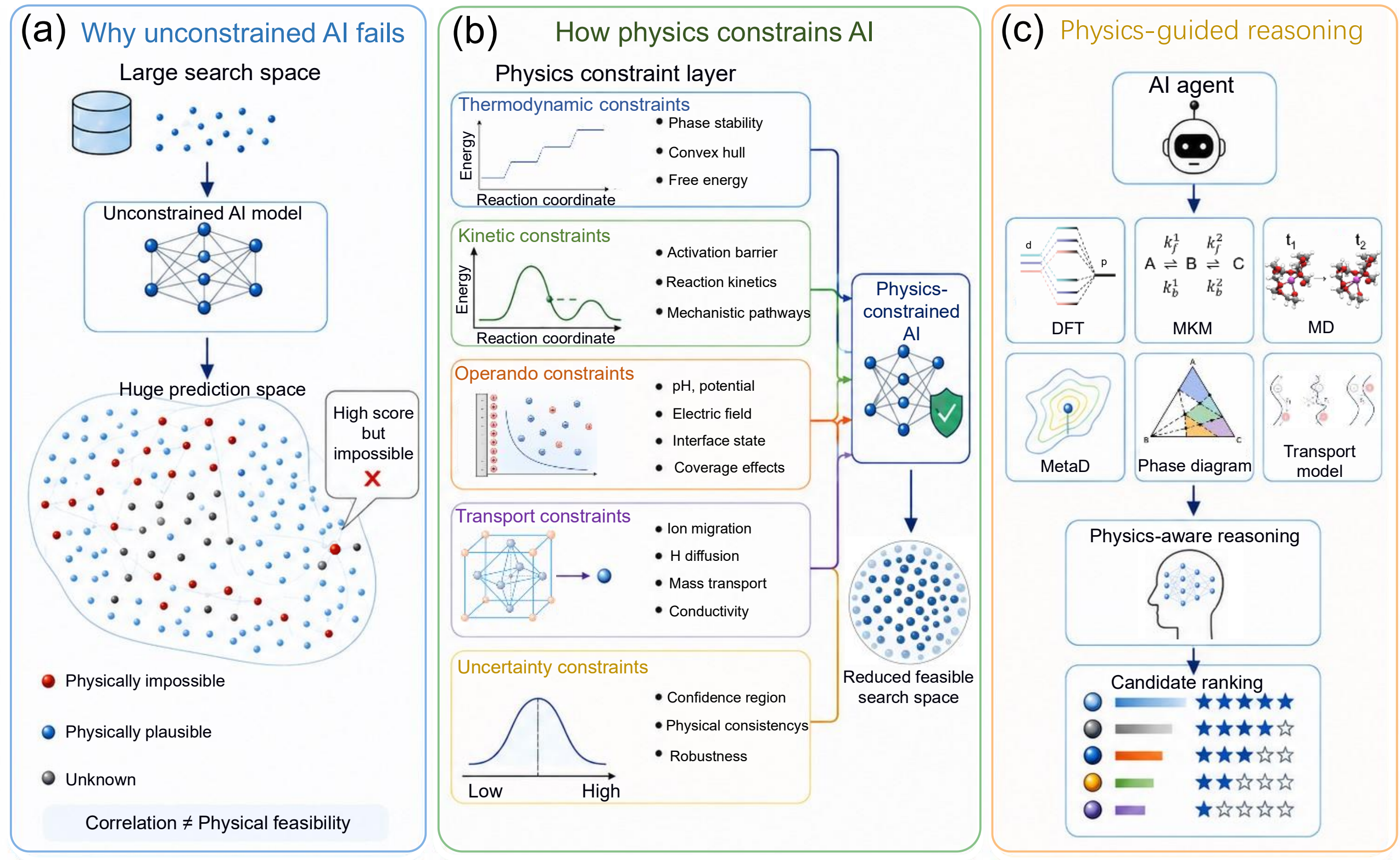

**Figure 5.** Conceptual illustration of physics as constraint in PhysMat AI. (a) Unconstrained AI may identify high-scoring candidates that violate fundamental physical principles. (b) Physics-based constraints, including thermodynamic stability, reaction kinetics, *operando* conditions, transport behavior, and uncertainty quantification, reduce the feasible search space and enforce physical consistency. (c) Emerging AI agents perform physics-guided reasoning by integrating first-principles calculations, kinetic modeling, phase stability analysis, and transport simulations, enabling reliable candidate evaluation and decision-making. MetaD: metadynamics simulations. MD snapshots and the transport-model illustration are reproduced with permission from ref.[92] licensed under a Creative Commons License CC BY-NC 3.0.

Based on these constraints, AI models or agents further evolve from a predictive tool into a physics-informed reasoning system (**Figure 5c**). Unlike conventional ML models that directly output predictions, emerging AI agents can invoke computational and theoretical tools, including DFT, microkinetic modeling (MKM), AIMD, enhanced-sampling techniques, phase-diagram analysis, and transport models, to evaluate candidate materials through iterative reasoning. **In this framework, physical constraints no longer serve merely as filters applied after prediction; instead, they become an integral part of the reasoning process itself.** As a result, AI performs mechanism-driven decision-making within a physically feasible search space, enabling a transition from correlation-driven AI to physics-guided reasoning AI.

The specific forms of these constraints vary across different materials systems. In catalysis, AI reasoning is primarily constrained by *operando* surface stability, reaction thermodynamics, kinetics, and electrochemical environments. For SSEs, constraints arise from ion transport, electrochemical stability, and interfacial compatibility. For hydrogen-storage materials, feasible operating windows are jointly determined by thermodynamics, diffusion kinetics, phase stability, and reversibility. The practical implementation of these constraints differs across catalysis, SSEs, and hydrogen storage. Representative examples are discussed below.

### 3.4 Physics as Verifier: External Physical Validation Beyond AI Prediction

The fourth core layer of PhysMat AI is that prediction does not end with AI output. Instead, predictions must undergo physical validation, namely physics outside AI. After AI generates candidate results, they must be further examined through DFT, kinetic modeling, diffusion simulations, and operando-condition validation.

This layer is a key component that distinguishes PhysMat AI from conventional data-driven models. Its primary objective is to establish a connection between AI predictions and the real physical world through both theoretical calculations and experimental validation. As illustrated in **Figure 6a**, PhysMat AI incorporates a verification toolbox composed of complementary theoretical and experimental approaches. Among them, DFT calculations provide fundamental information such as adsorption energies, electronic structures, and reaction barriers. Microkinetic analysis further links thermodynamic and kinetic processes, enabling the prediction of catalytic activity. Phase-stability analysis is employed to evaluate the thermodynamic feasibility of candidate materials,[93] while interface simulations are used to describe complex phenomena including electrode-electrolyte interfaces,[94] charge transfer, and interfacial stability. At the same time, experimental techniques such as electron microscopy, spectroscopic characterization, electrochemical measurements, and pore-structure analysis provide direct evidence for validating theoretical predictions.

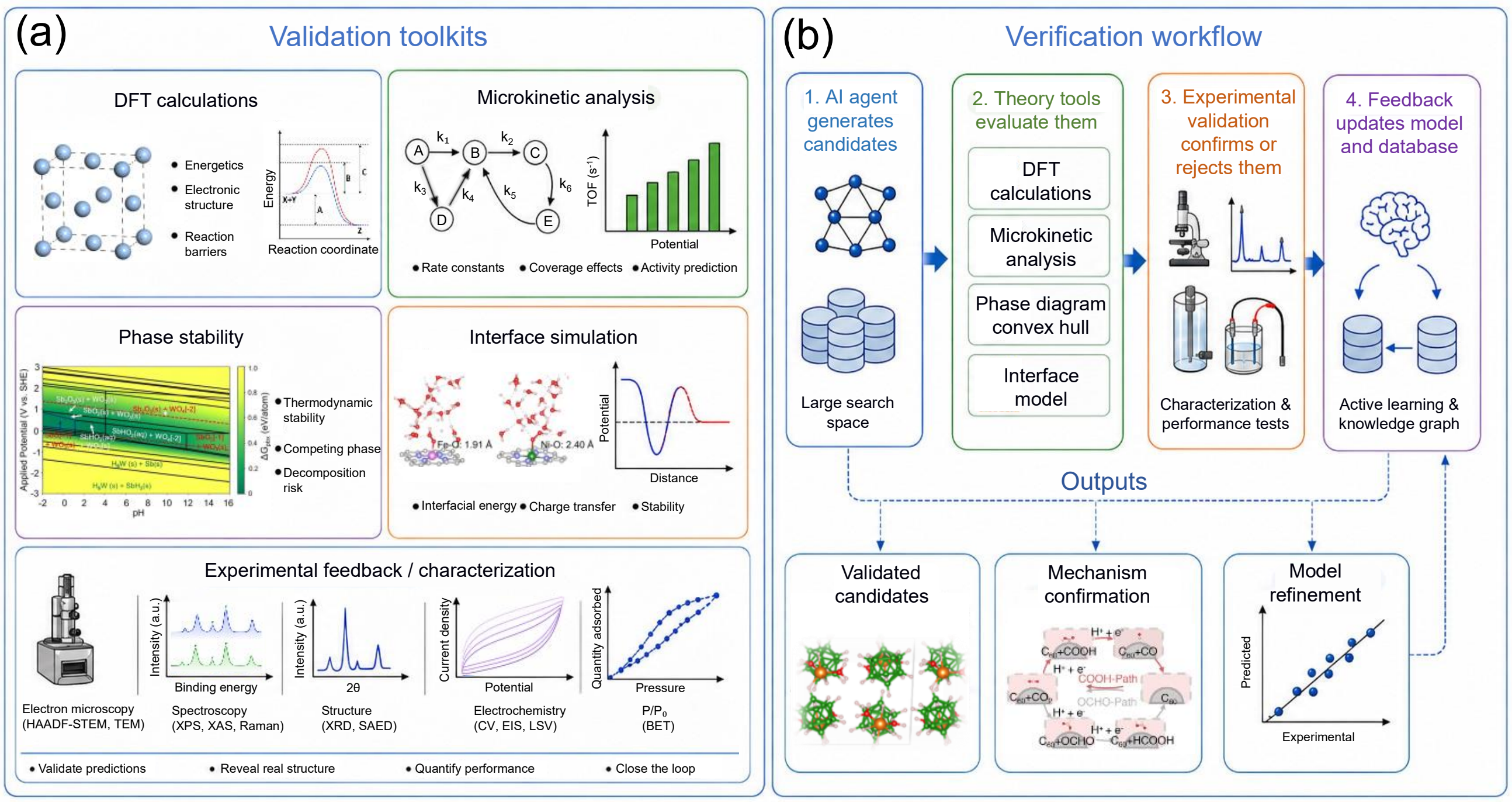

**Figure 6.** Physics as Verifier in PhysMat AI. (a) A validation toolkit combining theoretical calculations, simulations, and experimental characterization. Subpanel for phase stability is reproduced with permission from ref.[93] licensed under a Creative Commons License CC BY-NC 4.0. Subpanel for interface simulation is reproduced with permission from ref.[94] licensed under a Creative Commons License CC BY-NC 3.0. (b) A closed-loop verification workflow in which AI-generated candidates are evaluated by physics-based tools, validated experimentally, and fed back to continuously improve databases, knowledge graphs, and AI models. Subpanel for validated candidates is reproduced with permission from ref.[37] copyright 2023, American Chemical Society. Subpanel for mechanism confirmation is reproduced with permission from ref.[95] licensed under a Creative Commons License CC BY-NC 4.0.

### 3.4.1 General Description

Based on these tools, **Figure 6b** presents the verification workflow of PhysMat AI. First, AI agents can generate candidate materials from a large search space. These candidates are then screened and evaluated using theoretical methods, including DFT calculations, microkinetic modeling, phase-diagram analysis, and interface simulations. Promising candidates are subsequently subjected to experimental characterization and performance testing to confirm the validity and reliability of the predictions. Finally, the verified results are fed back into databases and knowledge graphs, enabling active learning and continuous model refinement within the PhysMat AI framework. Importantly, feedback should include not only successful validations but also failed predictions, ambiguous observations, and negative experimental results. These outcomes provide valuable information about the boundaries of the feasible design space, improve uncertainty estimation, and enable AI models to progressively refine their physical reasoning through iterative learning.

This verification system not only identifies experimentally validated candidate materials but also enables mechanism verification and model calibration, thereby continuously improving predictive accuracy and extrapolation capability. By integrating theoretical calculations, experimental validation, and data feedback into a unified framework, this role transforms materials AI from a simple prediction tool into a scientific discovery system capable of self-correction and continuous learning, providing reliable support for autonomous materials discovery in catalysis, solid-state battery, hydrogen storage, and related fields.

### 3.4.2 Examples in Catalysis, SSEs, and Hydrogen Storage

*In catalysis*, DFT calculations constitute the first verification layer in PhysMat AI for catalysis by validating adsorption energetics, reaction free energies, electronic structures, and reaction barriers predicted by ML models. However, thermodynamically favorable structures are not necessarily the true active states under operating conditions. Therefore, *operando* verification is required to evaluate catalyst stability under realistic reaction environments, including the effects of potential, pH, electric field, adsorbate coverage, electrolyte interactions, and dynamic surface reconstruction.[96] Building upon these verified active structures, microkinetic modeling further connects elementary-step energetics with reaction rates, surface coverages, and product selectivity, enabling the identification of kinetically accessible pathways and rate-determining steps. Together, DFT, microkinetic modeling, and *operando* analysis provide a multi-level verification framework that bridges theoretical predictions and realistic catalytic performance.

Experiment remains the ultimate verification layer. Predictions generated by AI models and physics-based simulations must ultimately be tested against reality. For example, Liu *et al*.[70] proposed a closed-loop catalyst discovery workflow integrating databases, ML, microkinetic modeling, catalyst screening, and experimental validation. Notably, the experimentally measured activity of the top-ranked $LiScO_2$ catalyst agreed with theoretical predictions within 5%, confirming that experimental verification is essential not only for validating candidate materials but also for refining models and closing the learning loop. Overall, AI-generated candidates must be validated through catalyst synthesis, electrochemical testing and *operando* characterization before being incorporated into subsequent model refinement. Unsuccessful candidates are equally informative because they reveal missing mechanisms or unrealistic assumptions, thereby improving subsequent model calibration and guiding future catalyst exploration.

*For SSEs in solid-state battery*, AI-generated hypotheses should be continuously tested through atomistic simulations, multiscale transport models, and experimental observations to establish reliable links between data-driven predictions and real electrochemical behavior. Although high-throughput screening and foundation models can rapidly generate candidate materials, validation remains indispensable because ion transport involves activated processes, rare events, local disorder, and interfacial phenomena that cannot be fully captured by prediction alone.

Universal MLIP frameworks can screen hundreds of thousands of structures, yet migration barriers may be systematically underestimated in activated or defective environments.[97] Benchmark platforms such as LiTraj have therefore been developed to evaluate ion migration performance directly rather than relying solely on force and energy errors.[98] **In this sense, physics-based verification is not a final formality, but a necessary part of the discovery process.**

Verification increasingly occurs through multiscale workflows that combine DFT, AIMD, metadynamics (MetaD), MLIP, and experiments. A representative example is the integration of SSE databases, LLMs, and *ab initio* MetaD for hydride electrolytes.[99] The workflow revealed molecular-unit-assisted transport, showing that conductivity trends alone are insufficient to identify the underlying migration pathway. Instead, MetaD was required to resolve local structural rearrangement and activated hopping processes. In MOF-based SSEs, LLM-assisted literature exploration and representation clustering enabled the identification of NOTT-400 as a candidate electrolyte, which was subsequently verified by physicochemical characterization and electrochemical testing.[100] These examples show that AI-assisted discovery becomes more convincing when prediction is coupled to material synthesis and battery-relevant validation. By integrating simulations, experimental characterization, and feedback, AI predictions can be continuously verified and refined, enabling more reliable and physically grounded discovery of next-generation SSEs.

*In hydrogen storage*, model confidence cannot be directly equated with physical credibility because key properties are strongly influenced by measurement conditions, sample history, catalyst distribution, phase constitution, and cycling state.[42, 63-67] Therefore, AI predictions should be continuously evaluated against thermodynamic consistency, dynamic behavior, and experimental observations.

Verification begins with data consistency. Reported capacities should distinguish theoretical, reversible, and experimentally accessible values, while hydrogen content and PCT-derived thermodynamic parameters should be standardized across different definitions and measurement conditions. Such consistency checks are essential because small deviations in hydrogenation enthalpy or entropy may lead to large errors in the predicted operating window.[64-65]

Simulation-based verification provides the second layer. DFT, thermodynamic calculations, NEB, MetaD, AIMD, and MLIP can evaluate phase stability, reaction energetics, diffusion pathways, and structural evolution, while quantum-nuclear treatments may be required when hydrogen transport involves tunneling effects.[66, 88-90, 101] These approaches verify whether the mechanisms inferred by AI are physically plausible rather than merely statistically correlated.

The final layer is experimental and *operando* verification. Techniques such as PCT measurements, temperature-programmed desorption (TPD), cycling tests, diffraction, and spectroscopy directly examine thermodynamic and kinetic behavior under realistic operating conditions. A notable example is the development of the physical model of $MgH_2$-based materials' hydrogen desorption,[102] where out-of-sample computational dehydrogenation barriers and experimental onset temperature were both used to validate the model accuracy as the final stage. In addition, we note that experimental feedback, including failed and negative results, should be incorporated into the data infrastructure to calibrate model uncertainty and continuously improve AI models. Despite differences in application scenarios, this verification philosophy is shared across catalysis, batteries, and hydrogen storage, providing a common route from AI prediction to physically reliable materials discovery.

### 3.5 Physics as Infrastructure: Toward Autonomous PhysMat AI Ecosystems

#### 3.5.1 General Description

The final layer of PhysMat AI is the construction of a physics-aware digital materials ecosystem, in which databases are no longer merely repositories of data but evolve into AI training platforms, infrastructures for mechanism reasoning, and ecosystems for autonomous discovery.[3, 103] As illustrated in **Figure 7a**, this data infrastructure integrates multiple domain-specific databases, including materials databases such as *DigMat*,[3] catalysis databases such as *DigCat* and its derivatives,[104-106] Open Catalyst,[107-108] battery databases such as *DigBat*/*DDSE*,[109] OEDB,[110] and hydrogen-storage databases such as *DigHyd*,[111-112] HydPark,[113] as well as widely used resources such as Material Project[114] and AFLOW.[115] The infrastructure can be further extended to other domains, including polymers, semiconductors, metal oxides, and two-dimensional materials.[116-118]

Unlike conventional databases that primarily store structural or property data, this infrastructure integrates experimental measurements, theoretical calculations, literature-derived information, multimedia data, as well as metadata and provenance records into a unified framework. Through the interconnection of these heterogeneous data sources, a rich and linked knowledge network is established, providing a common data foundation for materials AI and enabling more reliable learning, reasoning, and discovery. Knowledge graphs provide an effective semantic layer for this infrastructure by linking heterogeneous datasets through shared concepts, relationships, and standardized metadata. Such semantic integration enables AI agents to connect catalytic performance with synthesis conditions, characterization results, theoretical calculations, and operating environments, thereby supporting more reliable reasoning across diverse data sources.

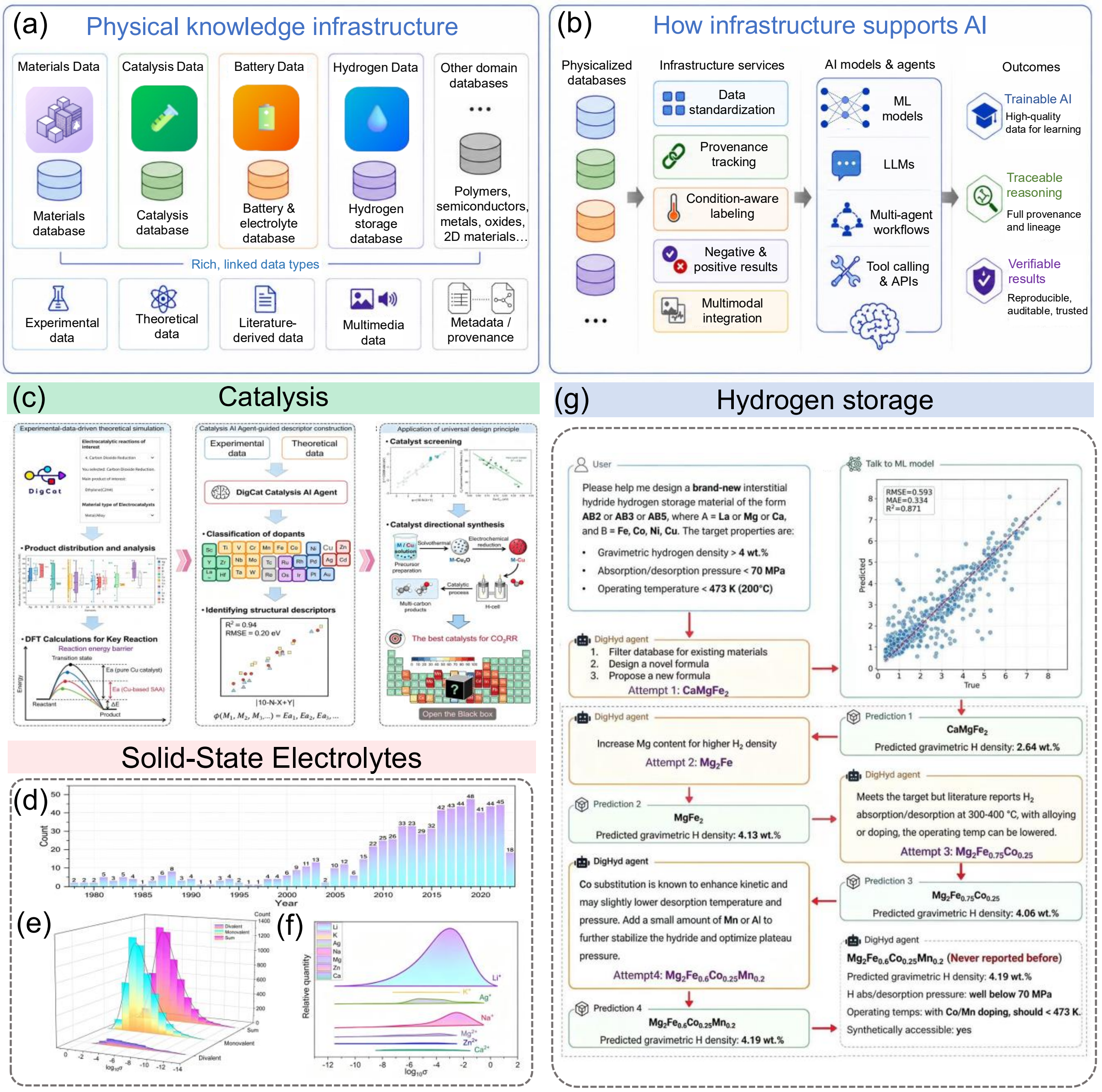

**Figure 7.** Physics as infrastructure in PhysMat AI. (a) A physical knowledge infrastructure built upon domain-specific databases and linked data resources. (b) Infrastructure services support AI models and agents through standardized, traceable, condition-aware, and multimodal data management, enabling reliable and verifiable materials intelligence. (c) *DigCat*-enabled AI workflow for catalyst screening, descriptor discovery, and catalyst design. Reproduced with permission from ref.[119] licensed under a Creative Commons License CC BY-NC 4.0. (d) Year-wise distribution of materials included in *DigBat* (formerly *DDSE*), updated to May 2023. (e) Conductivity distribution of monovalent and divalent solid

electrolytes contained in the database. (f) Cation-conductivity statistics of SSEs compiled from the database (unit: S/cm). Reproduced with permission from ref.[109] licensed under a Creative Commons License CC BY-NC 4.0. (g) *DigHyd*-powered inverse design of hydrogen-storage materials through AI-agent-guided optimization. Reproduced with permission from ref.[112] licensed under a Creative Commons License CC BY-NC 3.0.

Building upon this foundation, **Figure 7b** illustrates how the infrastructure supports AI systems. First, domain-specific databases are standardized and transformed into machine-readable physicochemical data resources. Subsequently, infrastructure services such as data standardization, provenance tracking, condition-aware labeling, positive and negative result management, and multimodal data integration are employed to ensure data quality and organize knowledge in a structured manner. Knowledge graphs further organize these standardized resources into machine-interpretable relationships, allowing AI agents to retrieve evidence across experiments, simulations, and literature rather than relying on isolated datasets. The processed data then support the training and deployment of ML models, LLMs,[120] intelligent agent systems,[121-124] as well as tool-calling frameworks.

As a result, the infrastructure not only provides high-quality training data but also records complete data provenance and reasoning processes, enabling traceable reasoning and verifiable outcomes. This layer should therefore be viewed not merely as a data repository, but as a semantically connected knowledge infrastructure in which standardized metadata, provenance records, ontologies, and knowledge graphs enable AI agents to perform reliable, traceable, and context-aware scientific reasoning.

### 3.5.2 Examples in Catalysis, SSEs, and Hydrogen Storage

*In catalysis*, the *DigCat* database provides the data foundation for AI-driven catalyst design. For example, by integrating experimental data with theoretical calculations, the Catalysis AI Agent can automatically identify correlations between catalytic performance and structural features and further construct physically meaningful descriptors, thereby uncovering general design principles governing the $CO_2$ reduction performance of Cu-based single-atom alloy catalysts (**Figure 7c**).[119, 125] This capability can be further enhanced by knowledge-graph-based organization, which semantically links catalyst composition, synthesis protocols, reaction conditions, and catalytic performance into a unified reasoning framework. Based on these design principles, high-performance catalysts can be rapidly screened and rationally synthesized. This work demonstrates that large-scale catalysis databases can transform fragmented experimental knowledge into learnable design rules, thereby accelerating the establishment of structure-performance relationships.

Other platforms further expand this concept. For example, Chen *et.al* developed the DigMethPy,[106] a digital catalysis platform specifically for the design of molten catalysts for methane pyrolysis reaction. By integrating literature knowledge, experimental data, computational results, ML models, and LLMs, DigMethPy establishes a closed-loop workflow encompassing data collection, modeling, prediction, validation, and feedback. This framework supports catalyst screening, knowledge organization, and continuous optimization, highlighting a pathway toward an autonomous PhysMat AI ecosystem built upon data-centric infrastructure. Furthermore, Jia *et al*.[126] developed StableOx-Cat Agent based on materials databases such as Materials Project, enabling the automated screening and intelligent discovery of metal oxide electrocatalysts with both thermodynamic and electrochemical stability. These examples collectively demonstrate how physics-aware data infrastructure can support AI agents in moving beyond data retrieval toward knowledge-driven catalyst discovery and design.

*For SSEs in solid-state battery*, conductivity depends not only on composition and structure but also on temperature, mobile ion species, sample density, processing history, electrode configuration, and measurement protocol. A single reported conductivity value therefore often conceals substantial physical information. This limitation becomes particularly important for AI agents, which require traceable and context-aware knowledge rather than isolated performance labels.

To address this challenge, dedicated electrolyte databases have emerged. The Dynamic Database of Solid-State Electrolyte (*DDSE*), which has recently evolved into the broader Digital Battery platform (*DigBat*, www.digbat.org), organizes electrolyte-specific information including ionic conductivity, activation energy, conducting ion, material family, and literature provenance.[109, 127] Rather than reducing transport behavior to a single room-temperature value, temperature-dependent conductivity measurements are preserved as independent records, allowing transport properties to remain linked to their experimental conditions.

As shown in **Figure 7d**, the continuous growth of the *DDSE* database provides an increasingly rich data foundation for AI-driven solid-state electrolyte discovery. At its initial release, *DDSE* integrated 678 performance records, covering over 600 SSE materials reported between 1978 and 2023. The database includes monovalent cations such as $Li^+$, $Na^+$, $K^+$, and $Ag^+$, as well as divalent cations including $Ca^{2+}$, $Mg^{2+}$, and $Zn^{2+}$, together with a wide range of anion chemistries such as halides, hydrides, sulfides, and oxides (**Figure 7e**). Moreover, the database continues to be actively updated. The collected data span a broad temperature range from 132.40 K to 1261.60 K and contain multidimensional information, including ionic conductivity, activation energy, phase-transition behavior, and literature provenance. As shown in **Figure 7f,** the conductivity distributions of monovalent and divalent SSEs reveal that more than 60% of the materials exhibit $\log_{10}$(S/cm) values between -5 and -2 $\log_{10}$(S/cm). Further analysis according to the conducting cation species shows that $Li^+$-based materials constitute the largest group, with conductivity values mainly concentrated between -4 and -2 $\log_{10}$(S/cm), whereas divalent-ion conductors generally exhibit lower ionic conductivity because of their stronger electrostatic interactions.

The current *DigBat* platform integrates 3725 experimental SSEs together with 846 computational SSE records containing activation energy information.[99] By aggregating $Li^+$, $Na^+$, $K^+$, $Ag^+$, $Ca^{2+}$, $Mg^{2+}$, and $Zn^{2+}$ conductors under consistent electrolyte-specific fields, *DigBat* allows systematic comparison across charge carriers, electrolyte families, and transport conditions. This organization is particularly important for agent-based discovery, because the agent must retrieve not only candidate materials but also the experimental or computational context in which each transport property was obtained.

Data-driven studies demonstrate the value of large-scale and structured electrolyte datasets for materials discovery. Large SSE datasets have been used to analyze relationships between ionic conductivity, activation energy, crystal chemistry, and transport mechanisms across multiple materials families.[127-128] Other analyses have quantified the exploration landscape of solid Li-ion conductors from the perspective of anion chemistry, revealing which chemical families have been extensively explored and which remain underrepresented.[129]

Building on these observations, structured SSE databases function as active knowledge systems rather than passive repositories. By integrating transport properties with synthesis conditions, measurement protocols, and computational results, they provide AI agents with context-aware knowledge for evidence retrieval, mechanistic reasoning, and hypothesis generation.

Instead of retrieving materials solely according to reported conductivity, AI agents can identify inconsistencies across datasets, detect knowledge gaps, and prioritize simulations or experiments that are expected to provide the highest information gain. The resulting hypotheses can then be evaluated through a mechanism-oriented workflow that integrates DFT, AIMD, MLIP, MetaD, and experimental validation. Rather than treating prediction and validation as separate steps, new evidence continuously refines descriptors, updates the knowledge base, and reprioritizes candidate materials, forming an iterative closed-loop discovery process. Because SSEs encompass diverse chemistries, transport mechanisms, and degradation pathways, future AI systems will likely adopt modular architectures that integrate specialized models through shared databases, interpretable descriptors, uncertainty estimation, and experimental feedback.

Collectively, standardized databases, provenance information, mechanism-aware descriptors, and continuous experimental feedback transform battery databases into active knowledge systems that support AI-driven reasoning and materials discovery. Such an ecosystem enables AI agents to retrieve context-aware knowledge, formulate mechanistic hypotheses, coordinate simulations with experiments, and continuously improve the discovery of next-generation SSEs.

*In hydrogen storage*, research has produced decades of PCT curves, TPD profiles, kinetic traces, cycling plots, diffraction refinements, atomistic calculations, and mechanistic interpretations. Much of this information is still embedded in figures or narrative text, making it difficult for AI systems to compare, verify, and reuse. Physics as infrastructure requires more than collecting values at scale. It requires transforming scattered literature into provenance-linked, thermodynamically coherent, descriptor-ready, and uncertainty-aware knowledge.[64-65, 112]

The multi-agent Descriptive Interpretation of Visual Expression (DIVE) workflow and *DigHyd* illustrate this transition. DIVE uses an AI-agent workflow to extract and organize graphical hydrogen-storage information, including PCT curves, TPD profiles, and related performance plots.[112] As shown in **Figure 7g**, *DigHyd* then provides a hydrogen storage specific platform in which compositions, capacities, enthalpies, entropies, equilibrium pressures, material classes, and literature provenance can be curated and analyzed.[65] The important point is that such platforms are not simply larger spreadsheets. They are physical memory systems: they preserve the conditions under which a value was measured, the thermodynamic interpretation used to derive it, and the source-level traceability needed to verify it.

A mature PhysMat AI ecosystem for hydrogen storage should connect this memory layer with descriptor models, simulation engines, physical constraints, and validation feedback. An AI agent could retrieve PCT-derived thermodynamic data, identify whether a candidate lies in a known capacity-pressure trade-off, select descriptor families appropriate to capacity or equilibrium pressure, request atomistic or mesoscale simulations, and prioritize experiments that reduce uncertainty. After validation, successful, failed, and ambiguous results should return to the database with metadata describing sample preparation, phase constitution, particle size, catalyst addition, activation protocol, gas purity, cycling history, and measurement uncertainty. Negative and null results are especially valuable because they define failure boundaries and improve calibration of future predictions.[64-65, 112, 130-131]

In this view, Japan's GteX hydrogen-storage program[132] and the emerging physics-aware AI ecosystem for hydrogen-storage materials[56] exemplify how research infrastructures are evolving from data repositories into integrated platforms for knowledge organization, AI reasoning, and autonomous materials discovery. Specifically, the database preserves knowledge; descriptors translate mechanisms into learnable representations; constraints keep reasoning physically feasible; verifiers confirm or reject predictions; and the infrastructure continuously updates the knowledge base through feedback. Hydrogen storage therefore illustrates why future materials AI should not merely predict isolated properties, but reason over physically constrained and experimentally verifiable operating windows. Together with catalysis and SSEs, this example further demonstrates that PhysMat AI provides a transferable framework in which common physical principles are embedded into AI through prior knowledge, descriptors, constraints, verification, and infrastructure.

## 4. Summary and Outlook

### 4.1 Overall Summary

Physics-guided AI is emerging as a new paradigm for materials discovery, moving beyond purely data-driven learning toward physically grounded intelligence. Throughout this *Perspective*, we highlighted the PhysMat AI framework, which integrates physical knowledge into AI systems through five complementary roles: **Physics as Prior, Descriptor, Constraint, Verifier, and Infrastructure**. Together, these layers establish a full-stack architecture that connects scientific knowledge, data resources, AI models, reasoning processes, and experimental validation. Most importantly, the five-layer PhysMat AI framework provides the foundation for the evolution from physics-aware AI to physics-autonomous AI.

Across catalysis, SSEs in solid-state battery, and hydrogen storage, we further demonstrate that these seemingly different materials systems are united by common physical principles, including thermodynamics, kinetics, energetics, and multiscale interactions, which enable the PhysMat AI framework to generalize across diverse applications. This shared physical foundation distinguishes PhysMat AI from conventional black-box AI, in which models primarily learn statistical correlations from data. Rather than treating AI as a black-box predictor, PhysMat AI embeds physical laws, mechanistic understanding, and verification workflows directly into the discovery process. By integrating physical principles with data-driven intelligence, PhysMat AI provides a pathway toward more reliable, explainable, and transferable materials discovery, bridging the gap between statistical prediction and scientific understanding.

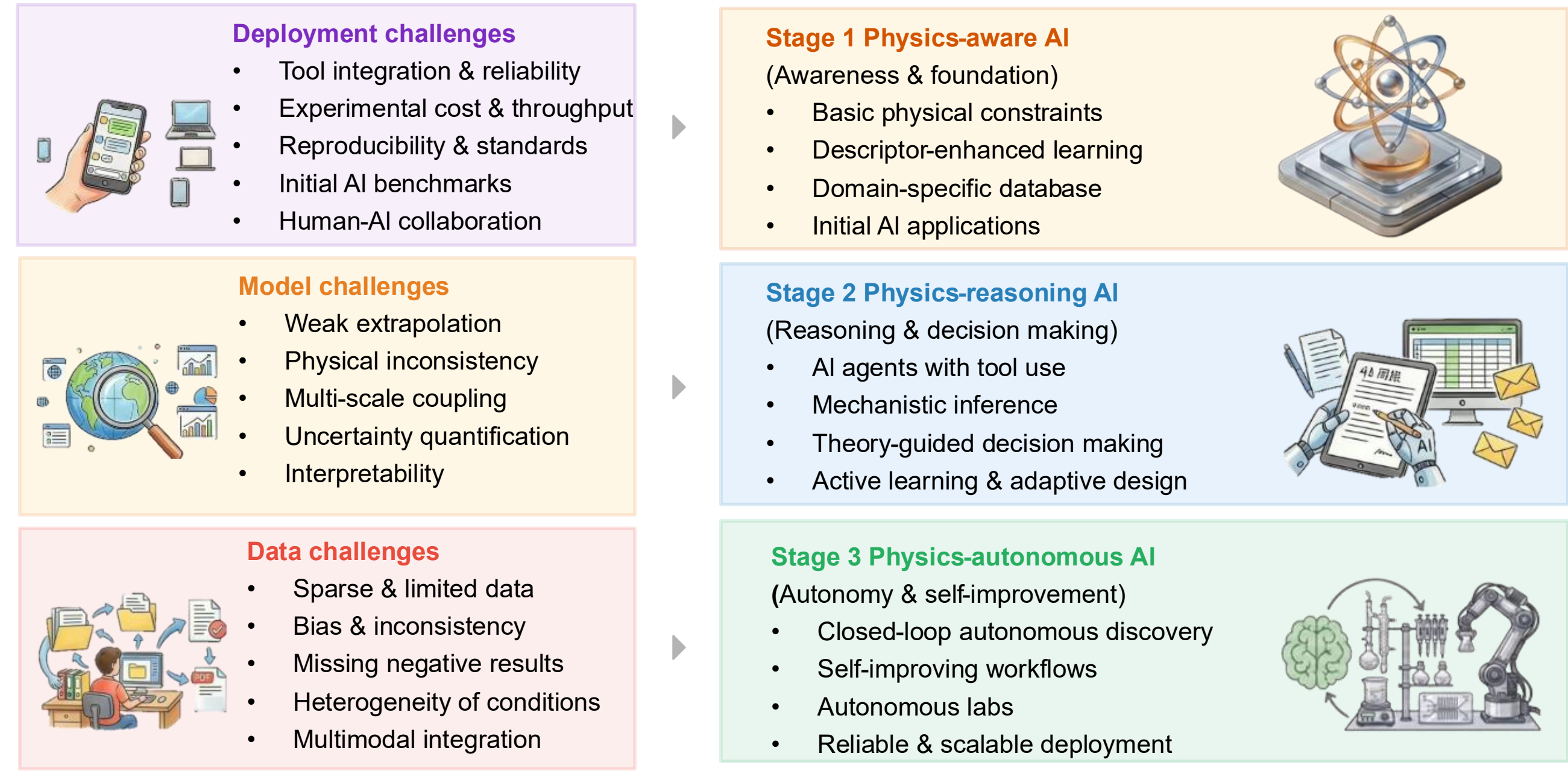

**Figure 8.** Challenges and roadmap of PhysMat AI. Key deployment, model, and data challenges facing PhysMat AI, together with a three-stage roadmap from physics-aware AI to physics-reasoning AI and ultimately physics-autonomous AI for reliable and autonomous materials discovery.

### 4.2 Challenges and Roadmap

Despite rapid progress, significant challenges remain before PhysMat AI can achieve fully autonomous scientific discovery. As illustrated in **Figure 8,** current limitations arise from three interconnected aspects: deployment challenges, model challenges, and data challenges. Practical deployment is still constrained by tool integration, experimental throughput, reproducibility, benchmarking standards, and effective human-AI collaboration. At the model level, weak extrapolation, physical inconsistency, uncertainty quantification, interpretability, and multiscale coupling remain major obstacles. Meanwhile, fragmented databases, limited negative results, heterogeneous experimental conditions, and insufficient multimodal integration continue to restrict the quality of AI training and evaluation.

Looking forward, the development of PhysMat AI may evolve through three progressive stages, each characterized by increasingly advanced reasoning capability, physical consistency, tool autonomy, and closed-loop learning. In **Stage 1** (Physics-aware AI), AI systems incorporate basic physical constraints, domain knowledge, descriptors, and curated databases, enabling learning beyond raw data correlations. At this stage, AI primarily improves prediction reliability within known materials spaces, while extrapolation capability and uncertainty awareness remain limited. In **Stage 2** (Physics-reasoning AI), AI agents and multi-agent systems become capable of invoking theoretical models and computational tools to perform mechanism-aware reasoning, theory-guided decision-making, and adaptive materials design. Beyond prediction, AI agents can formulate and evaluate mechanism-based hypotheses, invoke appropriate computational tools, and quantify confidence in competing explanations. Ultimately, **Stage 3** (Physics-autonomous AI) envisions self-improving, closed-loop discovery systems that integrate AI prediction, physical validation, theoretical calculations, and autonomous experimentation. Autonomous systems further integrate experimental feedback, continuously update their knowledge base, and iteratively improve model reliability under previously unseen conditions.

Overall, the ultimate goal of PhysMat AI is not merely to accelerate materials screening, but to establish a new scientific paradigm in which AI and physical science co-evolve, transforming materials research from data-driven prediction to physics-guided understanding and autonomous discovery.

## Notes

The authors declare no competing financial interest.

## ACKNOWLEDGMENT

The authors thank the support from JSPS KAKENHI (No. JP25H01508).